\documentclass[runningheads]{svmult}

\usepackage{makeidx}   
\usepackage{graphicx}  
\usepackage{subeqnar}  
\usepackage{multicol}  
\usepackage{physprbb}  
\makeindex             

\begin{document}
\title*{Physical Constraints From Broadband Afterglow
Fits: GRB000926 as an Example}
\toctitle{Physical Constraints From Broadband Afterglow Fits:GRB000926 as an Example}
%
%
\titlerunning{Broadband Afterglow Fits: GRB000926}
%

\author{Sarah A. Yost\inst{1,2}
\and Re'em Sari\inst{1}
\and Fiona A. Harrison\inst{1}
\and Edo Berger\inst{1}
\and Alan Diercks\inst{1}
\and Titus Galama\inst{1}
\and Dan Reichart\inst{1}
\and Dale Frail\inst{3}
\and Paul A. Price\inst{4}}
\authorrunning{S.A. Yost et al.}
%
%
\institute{Caltech, Pasadena CA 91125, USA
\and yost@srl.caltech.edu
\and National Radio Astronomy Observatory, Socorro NM 87801, USA
\and Research School of Astronomy \& Astrophysics, Australian National
University, ACT 2601, Australia}

\maketitle              

\begin{abstract}
We develop a model to fit the broadband afterglows of GRBs from the
intrinsic parameters of the fireball's synchrotron emission, and apply it
to a few well-studied events, with the goal of constraining the intrinsic
variability of GRB parameters. We give an example here of fitting to the
recent bright event GRB000926.
\end{abstract}

\section{Introduction}
The single successful model of GRB emission to date has been the fireball
model. A small amount of matter is accelerated to a large Lorentz factor
$\Gamma$. Shock expansion produces synchrotron emission of radiation with
a well-defined spectrum. The spectral breaks
$\nu_{break}$ are functions of fireball parameters and depend on the
hydrodynamics of the fireball's evolution. The hydrodynamics are strongly
affected by the
environment and geometry of the fireball, thus the afterglow's broadband
lightcurves can in principle constrain fundamental parameters of the
burst. For example, collimation of the ejecta produces an achromatic
break, but the evolution of $\nu_{break}$ past observed frequencies does
not.

We consider two possible density profiles for the burst
environment, $r^{0}$ as in the interstellar medium (ISM) and
$r^{-2}$ as from a constant stellar wind (WIND).

We calculate the synchrotron flux as a function of $t$,$\nu$ from the
luminosity
distance and redshift, and a set of fundamental parameters:
isotropic-equivalent energy $E$, electron powerlaw index $p$, electron and
magnetic energy fractions, $\varepsilon_{e}$ and $\varepsilon_{B}$,
as well as the circumburst density: a constant $n$ in the
ISM case or $A$ ($\rho = Ar^{-2}$) in the WIND case. The equations
are based on Sari et al~\cite{sari98} and Granot et 
al~\cite{gps99a},~\cite{gps99b} for the ISM model and Chevalier \&
Li~\cite{cheli00} for the WIND model. Collimation effects
on the evolution are based upon Sari et al~\cite{sari99}. We include the
effects of inverse Compton scattering based upon Sari 
\& Esin~\cite{sari00}.
Host extinction is parametrized by $A_{V}$ according to the prescription
of Reichart~\cite{reich}.

This calculated flux is compared to observations corrected
for Galactic extinction and host flux, and a Powell gradient search
optimizes the model parameters.

\section{GRB000926: Preliminary Results}

The IPN detected this event on 2000 September 26.993 and
rapidly disseminated its postion, leading to
observations by many. We use the optical observations at
$\le 1$ day post-burst by Hjorth et al~\cite{hjorth} and Fynbo et 
al~\cite{fynbo}, along with the data presented in Price et
al~\cite{paul}, with its calibration and host flux subtraction, as
well as the x-ray data of 
Piro \& Antonelli~\cite{piro00}. We allow a systematic
calibration uncertainty of 4\%, account for interstellar scintillations
in the radio based
on Walker~\cite{walker}, and calculate the best-fit ISM model (I) and WIND
model (W).

\begin{figure}[ht]
\unitlength1cm
\mbox{\small
\begin{minipage}[ht]{12cm}
 \begin{minipage}[t]{6cm}
  \mbox{}\\ 
  \begin{picture}(6,5.5)
  \includegraphics[width=\textwidth]{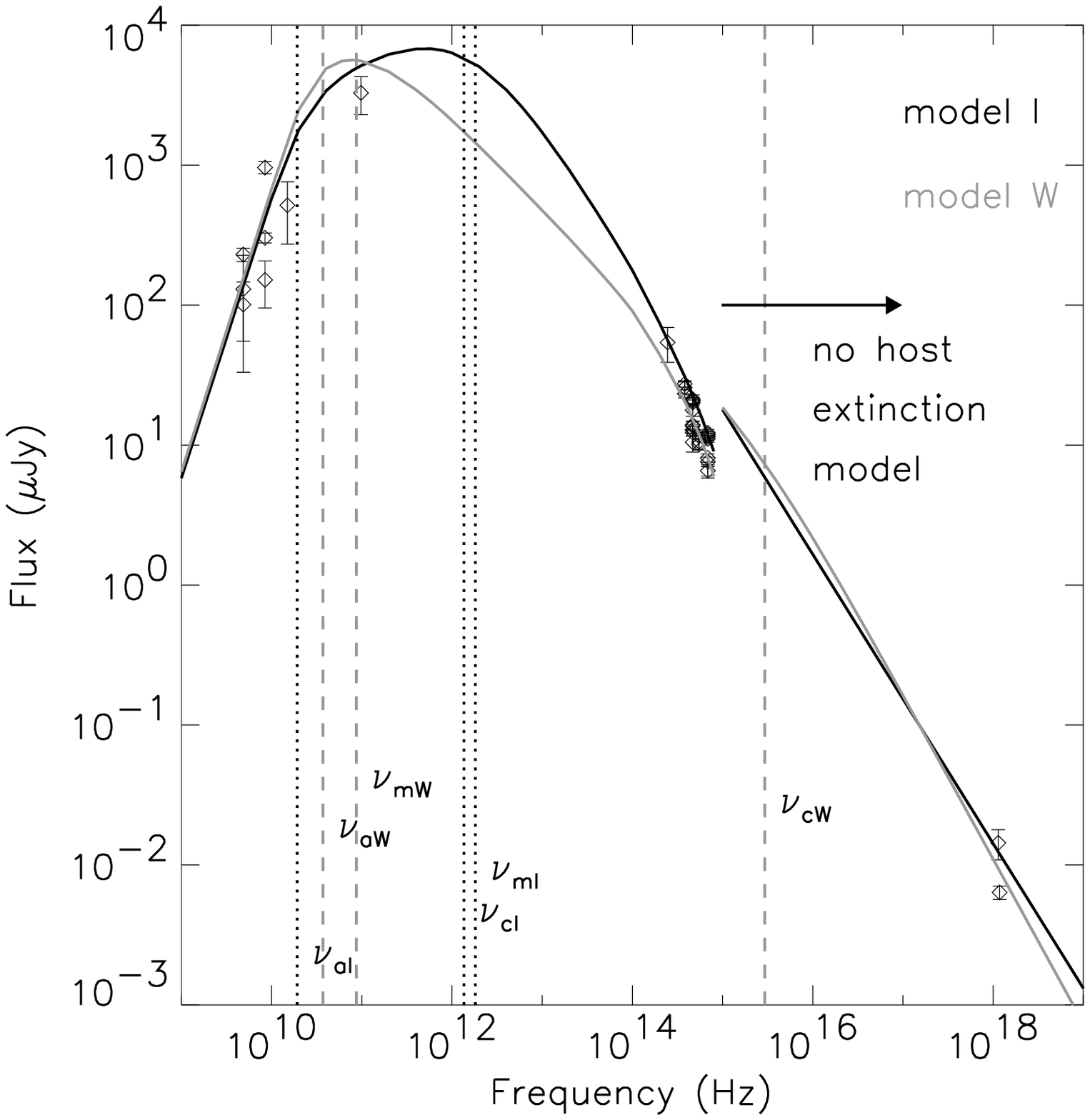}
  \end{picture}
  \label{eps2}
  \par
\caption[]{}
 \end{minipage}
\hfill
\parbox[t]{6cm}{\makebox[0cm]{}\\{Broadband spectra of model I (black)
and W (grey) at 2 days. Data from $2\pm1$ days is plotted over the curves,
interpolated to day 2 by model I. Both models provide a reasonable fit to
the broadband data.

  Inverse Compton (IC) cooling constrains the relative evolution of
$\nu_{break}$, preventing a high $\nu_{c}$ to better fit the x-ray.
A fit to the ISM with no IC gives a notably different fit, with a much
higher $\nu_{c}$. IC effects are not trivial and must be included in model
fits.} }
\end{minipage} }
\end{figure}
\begin{table}
\caption{Model Parameters}
\begin{center}
\renewcommand{\arraystretch}{1.0}
\setlength\tabcolsep{5pt}
\begin{tabular}{llll|llll}
\hline\noalign{\smallskip}
  & I & W & units & & I & W & units \\
\noalign{\smallskip}
\hline
\noalign{\smallskip}
$E$ & 1.1 & 39 & $10^{53}$erg & $p$ & 2.1 & 2.2 &  \\
$n$ & 0.62 &   & $cm^{-3}$  & $A$ &   & 1.5 & $5\times10^{11} gcm^{-2}$ \\
$\varepsilon_{e}$ & 0.27 & 0.012 &  & $\varepsilon_{B}$ & 
0.95 & 0.0025 &  \\
$\theta$ & 0.083 & 0.044 & rad & $t_{jet}$ & 1.2 & 1.8 & day\\
$A_{V}$ & 0.2 & 0.3 & mag \\
\hline
\end{tabular}
\end{center}
\label{Tab1a}
\end{table}

The fit to model I, including radio scintillation effects, gives a total
$\chi^{2}$ of 197 for 80 degrees of freedom. Model W has $\chi^{2}=171$ for 80
d.o.f.. Both models assume an LMC-like host extinction curve,
though an SMC-like curve gave scant difference in the results. 

Models I and W both give a reasonable description of the data. Model W
gives a better optical fit, but does not seem to fit the late-time radio
data. Very late radio data may provide the best discriminant between the
ISM and WIND.

\begin{figure}[ht]
\unitlength1cm
\begin{minipage}[ht]{6.0cm}
\begin{picture}(6.0,6.0)
\includegraphics[width=\textwidth]{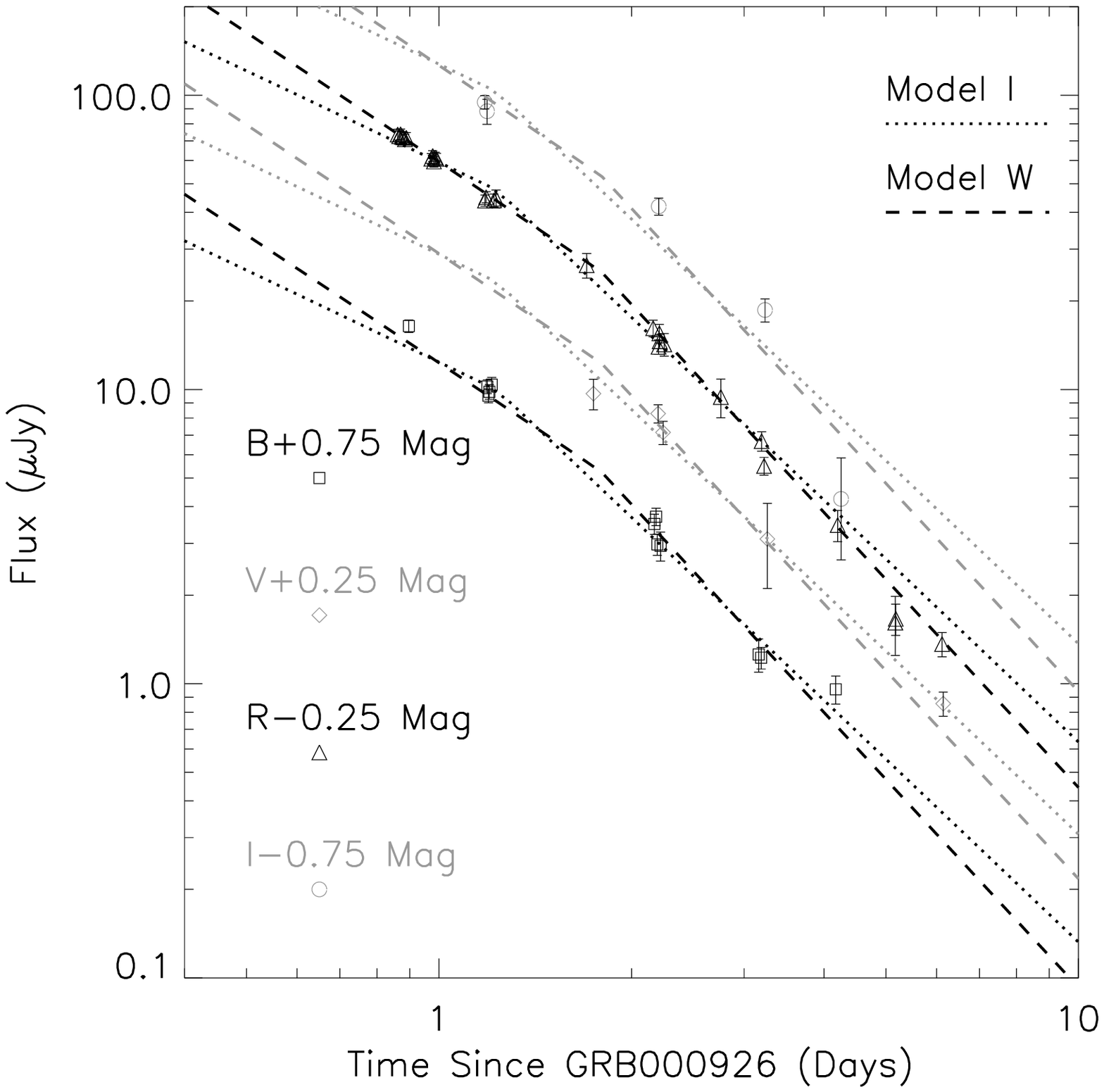}
\end{picture}
\par
\caption[]
{GRB000926 optical lightcurves. Both I and W models fit
reasonably well, though the R band is better fit by W, with its later jet
break and steeper slope.}
\end{minipage}
\hfill
\begin{minipage}[ht]{6.0cm}
\begin{picture}(6.0,6.0)
\includegraphics[width=\textwidth]{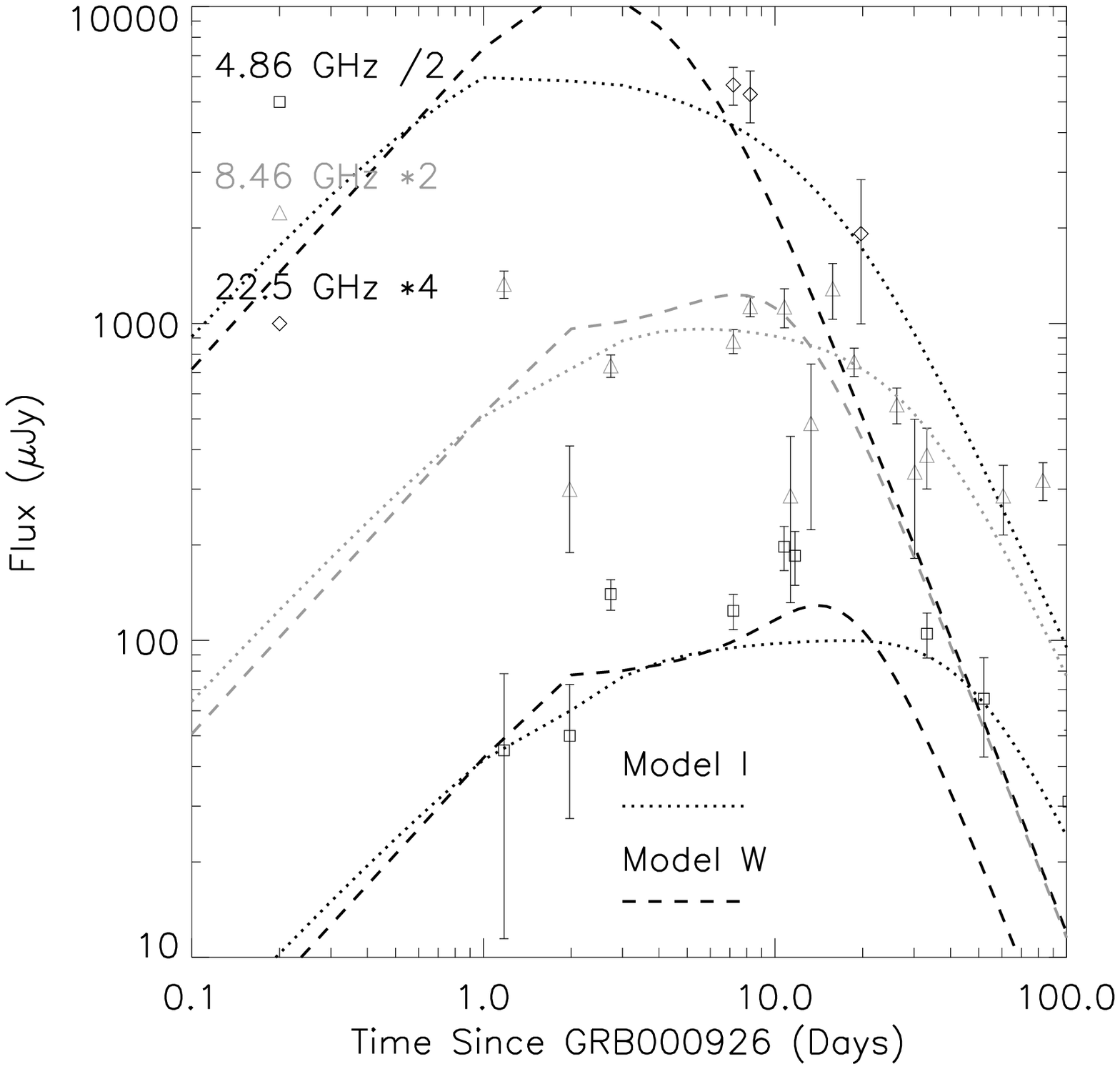}
\end{picture}
\par
\caption[]
{GRB000926 Radio lightcurves. At 22.5 GHz, early observations
could distinguish the models. At lower $\nu$, I and W are distinguishable
at late times, and I appears to fit better after about a month.}
\label{eps4}
\end{minipage}
\end{figure}

\section{Conclusion} ISM and WIND models fit the afterglow of GRB000926,
with non-negligeable Inverse Compton effects. The WIND underpredicts the
late 8.46 GHz data, whereas the ISM model is a far better fit to the late
radio observations, providing some evidence that this burst occured in a
medium of constant density.

%


\begin{thebibliography}{8.}
\addcontentsline{toc}{section}{References}

\bibitem{cheli00} Chevalier, R.A. and Li, Z. ApJ \textbf{536}, 195 (2000)

\bibitem{fynbo} Fynbo, J.P.U. et al. GCN 825 (2000)

\bibitem{gps99a} Granot, J., Piran, T. and Sari, R. ApJ \textbf{513}, 679
(1999a) 
\bibitem{gps99b} Granot, J., Piran, T. and Sari, R. ApJ \textbf{527}, 236
(1999b)

\bibitem{hjorth} Hjorth, J. et al GCN 809 (2000)

\bibitem{piro00} Piro, L. and Antonelli, L.A. GCN 832,833 (2000) 

\bibitem{paul} Price, P.A. et al. ApJL accepted, astro-ph 0012303 (2000)

\bibitem{reich} Reichart, D. astro-ph 9912368 (1999)

\bibitem{sari00} Sari, R. and Esin, A. A. astro-ph 0005253 (2000)

\bibitem{sari99} Sari, R., Piran, T. and Halpern, J.R. 1999 ApJ
\textbf{519} L17

\bibitem{sari98} Sari, R, Piran, T. and Narayan, R. 1998 ApJ \textbf{497}
L17

\bibitem{walker} Walker, M. A. MNRAS \textbf{294}, 307 (1998)

\end{thebibliography}
\end{document}